\documentstyle[12pt]{article}
\input epsf

\begin{document}

\begin{titlepage}

\begin{flushright}
CERN-TH/98-246\\
UM-TH 98-13\\
LAPTH 693-98\\
July 1998
\end{flushright}
\vspace{1.cm}

\begin{center}
\large\bf
{\LARGE\bf Nonperturbative three-point functions of the $O(N)$ 
           sigma model in the $1/N$ expansion at NLO}\\[2cm]
\rm
{Adrian Ghinculov$^{a,b,}$\footnote{Work supported by the 
                                  US Department of Energy (DOE)} 
 and Thomas Binoth$^c$}\\[.5cm]

{\em $^a$Randall Laboratory of Physics, University of Michigan,}\\
      {\em Ann Arbor, Michigan 48109--1120, USA}\\[.2cm]
{\em $^b$CERN, 1211 Geneva 23, Switzerland}\\[.2cm]
{\em $^c$Laboratoire d'Annecy-Le-Vieux de Physique 
         Th\'eorique\footnote{URA 1436 associ\'ee \`a l'Universit\'e de Savoie.} LAPP,}\\
      {\em Chemin de Bellevue, B.P. 110, F-74941, 
           Annecy-le-Vieux, France}\\[3.cm]
      
\end{center}
\normalsize

\begin{abstract}
We present a calculation of the three-point functions
of the $O(N)$-symmetric sigma model. The calculation is done
nonperturbatively by means of a higher-order $1/N$ expansion combined
with a tachyonic regularization which we proposed in previous publications.
We use the results for calculating the standard model process 
$f\bar{f} \rightarrow H \rightarrow WW$ nonperturbatively in
the quartic coupling of the scalar sector.
\end{abstract}

\vspace{3cm}

\end{titlepage}


\title{Nonperturbative three-point functions of the $O(N)$
       sigma model in the $1/N$ expansion at NLO}

\author{Adrian Ghinculov$^{a,b,}$\footnote{Work supported by the 
                                         US Department of Energy (DOE)} 
         and Thomas Binoth$^c$}

\date{{\em $^a$Randall Laboratory of Physics, University of Michigan,}\\
      {\em Ann Arbor, Michigan 48109--1120, USA}\\[.2cm]
      {\em $^b$CERN, 1211 Geneva, Switzerland}\\[.2cm]
      {\em $^c$Laboratoire d'Annecy-Le-Vieux de Physique 
         Th\'eorique\thanks{URA 1436 associ\'ee \`a l'Universit\'e de Savoie.} LAPP,}\\
      {\em Chemin de Bellevue, B.P. 110, F-74941, 
           Annecy-le-Vieux, France}}

\maketitle

\begin{abstract}
We present a calculation of the three-point functions
of the $O(N)$-symmetric sigma model. The calculation is done
nonperturbatively by means of a higher-order $1/N$ expansion combined
with a tachyonic regularization which we proposed in previous publications.
We use the results for calculating the standard model process 
$f\bar{f} \rightarrow H \rightarrow WW$ nonperturbatively in
the quartic coupling of the scalar sector.
\end{abstract}


In previous publications \cite{oneovn} 
we developed a nonperturbative approach for calculating processes
where the quartic self-coupling of the scalar sector of the
standard model becomes large, and therefore usual perturbation theory
becomes unreliable. This nonperturbative approach is based
on extending the standard Higgs sector to an $O(N)$-symmetric
sigma model, calculating the scattering amplitudes nonperturbatively
as a power series in $1/N$, and recovering the standard model in the
limit $N=4$. The connection to the physics of electroweak vector 
bosons is provided by the equivalence theorem. 

The idea of expanding in the number of degrees of freedom of 
the theory under consideration instead of the coupling constant
is rather old \cite{coleman,other1ovn}. 
This is in principle a very attractive idea
which attempts to find a solution which is valid at strong
coupling as well as at weak coupling, and which is free
of the renormalization scheme ambiguities associated
with conventional perturbation theory.

However, perturbation theory was by far wider used than 
the $1/N$ expansion for phenomenological purposes. 
An enormous amount of work in the recent years resulted in powerful
tools for the computation of higher-loop Feynman graphs.
The success of perturbation theory is due to the fact
that Feynman diagrams can be calculated for any theory,
and --- with increasing difficulty though --- in higher loop-orders.

These are precisely the issues which limited the applicability
of the nonperturbative methods based on the $1/N$ expansion.
First, the precise structure of the $1/N$ coefficients depends
on the theory under consideration. 
This structure can become so complicated as to
make a direct computation of all the graphs prohibitive even for the first
term of the $1/N$ expansion. A classical example of this type 
is the planar QCD in four dimensions \cite{thooft},
where the topological structure of the graphs is known, but
they could not be actually calculated so far. Second, in most
cases of physical interest the actual value of $N$ is not that
large as to make the leading order a good approximation. 
Such an example is the Higgs physics with which we deal in this letter.
Another example is the ordinary QED treated as the $N_e=1$ limit of an Abelian
theory coupled to $N_e$ species of electrons. 
When $N$ is not large enough, the leading
order solution in $1/N$, although simple, is numerically 
not an approximation of acceptable accuracy for phenomenological purposes.
Thus the inclusion of higher-order corrections is mandatory.
Higher-order contributions in $1/N$ are however difficult to calculate.
This is because of technical problems of combinatorial nature, and 
also because of the leak of techniques to calculate certain
classes of multiloop graphs. On the fundamental side there is
the question of treating renormalon-type chains inside higher-order
$1/N$ graphs.

For the case of the $O(N)$-symmetric sigma model, we 
have shown that the problems enumerated above 
can be dealt with, and higher-order
calculations in the $1/N$ expansion are feasible \cite{oneovn,forthcoming}.
The problem of casting all higher-loop graphs into a manageable
form is solved by using the auxiliary field formalism due to 
Coleman, Jackiw and Politzer \cite{coleman}. The tachyon problem
is dealt with by means of a minimal tachyonic subtraction
\cite{oneovn}.
The problem of evaluating the necessary multiloop graphs 
is solved numerically, by adapting a numerical three-loop 
technique for massive diagrams \cite{oneovn,3loop}.

The two-point functions of the $O(N)$ sigma model 
were calculated in refs. \cite{oneovn}.
This leads to a nonperturbative relation between the Higgs mass and 
width at next-to-leading order in $1/N$. Since in the scalar sector
of the standard model perturbation theory has already been extended
at two-loop level \cite{calc2loop,riesselmann,jikia}, 
this allows a strong test of the 
nonperturbative $1/N$ result at weak coupling. Indeed,
at weak coupling there is an impressive numerical agreement
between two-loop perturbation theory and the next-to-leading order
$1/N$ expansion. It is the purpose of this letter to extend
these results for the three-point functions of this theory.

As in the case of two-loop functions, we start with the
ordinary Lagrangian of an $O(N)$-symmetric sigma model.
Following ref. \cite{coleman}, we modify this Lagrangian by introducing 
a non-dynamical auxiliary field $\chi$:

\begin{eqnarray}
  {\cal L}   & = &
               \frac{1}{2}           
               \partial_{\nu}\Phi_0 \partial^{\nu}\Phi_0 
             - \frac{\mu_0^2}{2}      \Phi_0^2 
	     - \frac{\lambda_0}{4! N} \Phi_0^4 
             + \frac {3 N}{2 \lambda_0} 
                 (\chi_0 - \frac{\lambda_0}{6 N} \Phi_0^2 - \mu_0^2)^2 
	\nonumber \\	
           & \equiv &				     
    \frac{1}{2} \partial_{\nu}\Phi_0 \partial^{\nu}\Phi_0 
  - \frac{1}{2} \chi_0 \Phi_0^2 
  + \frac{3 N}{2 \lambda_0} \chi_0^2
  - \frac{3 \mu_0^2 N}{\lambda_0} \chi_0 
 ~~ , ~~ \Phi_0 \equiv \left( \phi_0^1, \phi_0^2, \dots , \phi_0^N \right)
\end{eqnarray}

This does not change the physical content of the sigma
model, because if one eliminates $\chi$ by using its equation
of motion, one recovers the original Lagrangian. 
The advantage of the auxiliary field formalism is that 
the quartic coupling of
the $\Phi$ field disappears, being replaced by trilinear 
vertices which involve the $\chi$ field. This results
into an enormous simplification of the possible topologies
of multiloop diagrams which may appear in higher orders of $1/N$.

The multiloop Feynman diagrams which contribute to the 
three-point functions of this theory at NLO in $1/N$
are shown in figure 1. Actually we define in this picture
the subtracted $1/N$ graphs $\hat{E}_1$, $\hat{F}_1$,  
$\hat{F}_2$ and $\hat{F}_3$. These are the ultraviolet finite
combinations which actually appear in the expressions of
observable physical quantities upon inclusion of the necessary
$1/N$ counterterms.

\begin{figure}[t]
\begin{tabular}{cccccccc}
      $i \frac{v}{\sqrt{N}} \hat{E}_1(s)$
    & $=$
    & 
    &       \raisebox{-.74cm}{ \epsfxsize = 3.3cm \epsffile{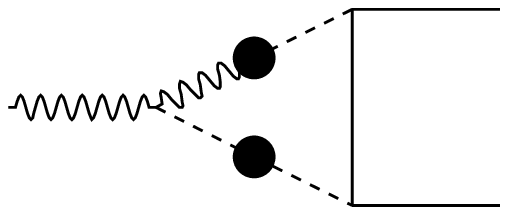} }  
    & $-$ & \raisebox{-.74cm}{ \epsfxsize = 3.3cm \epsffile{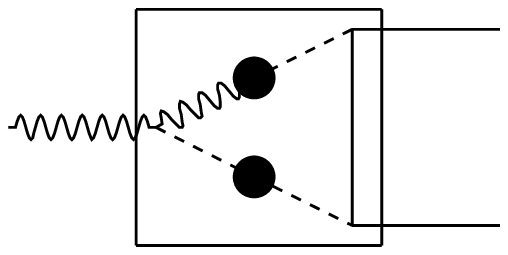} }  
    & 
    & 
  \\
      $i \frac{v}{\sqrt{N}} \hat{F}_1(s)$
    & $=$
    & 
    &       \raisebox{-.74cm}{ \epsfxsize = 3.3cm \epsffile{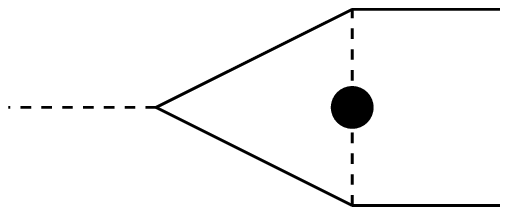} }  
    & $-$ & \raisebox{-.74cm}{ \epsfxsize = 3.3cm \epsffile{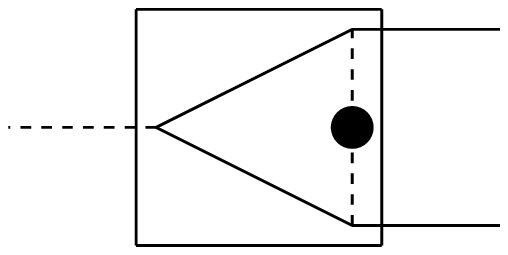} }  
    & 
    & 
  \\
      $i \frac{v}{\sqrt{N}} \hat{F}_2(s)$
    & $=$
    & 
    &       \raisebox{-.74cm}{ \epsfxsize = 3.3cm \epsffile{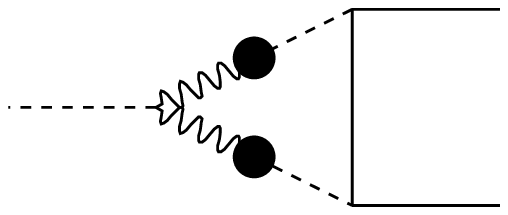} }  
    & $-$ & \raisebox{-.74cm}{ \epsfxsize = 3.3cm \epsffile{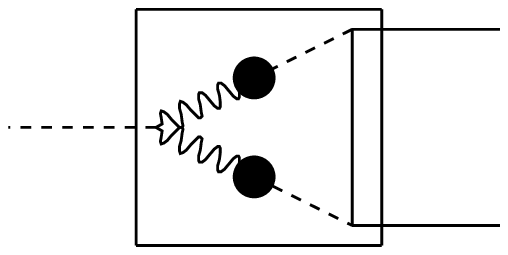} }  
    & 
    & 
  \\
      $i \frac{v}{\sqrt{N}} \hat{F}_3(s)$
    & $=$
    & 
    &       \raisebox{-.74cm}{ \epsfxsize = 3.3cm \epsffile{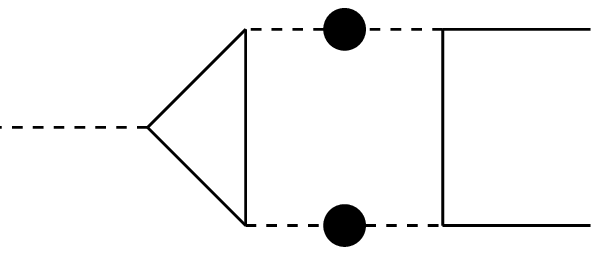} }  
    & $-$ & \raisebox{-.74cm}{ \epsfxsize = 3.3cm \epsffile{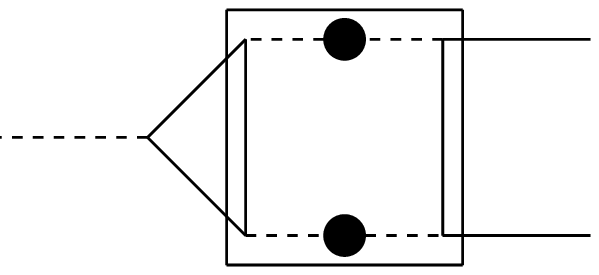} }  
    & 
    & 
  \\
    &       
    & $-$ & \raisebox{-.74cm}{ \epsfxsize = 3.3cm \epsffile{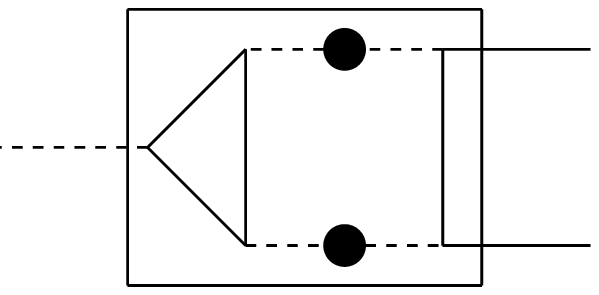} } 
    & $+$ & \raisebox{-.74cm}{ \epsfxsize = 3.3cm \epsffile{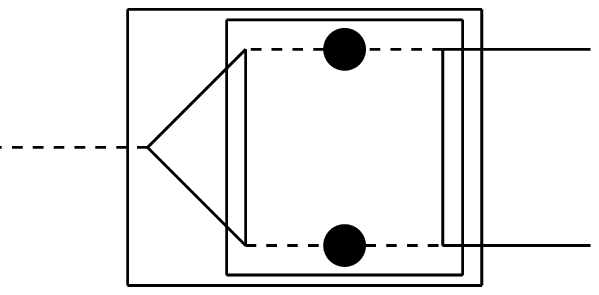} } 
    & 
    &  
\end{tabular}
\caption{{\em The definition of the subtracted vertex graphs. The blob
              on internal lines indicates the dressed nonperturbative
              propagators at LO in $1/N$, which result upon 
              summation of infinite chains of one-loop bubble diagrams.
              The wavy line is the $\sigma$ field,
              the solid line is the $\pi$ field, and the dashed line is
              the $\chi$ field.
              The boxes indicate internal and overall ultraviolet 
              subtractions performed at an arbitrary 
              subtraction scale $\mu$.}}
\end{figure}

The subtracted three-point graphs of figure 1 are calculated 
numerically with the same methods as the two-point graphs 
\cite{oneovn}.
The main difference is that, because of the different kinematic
combination of the external momenta, the rotation of the spatial
component of the loop momentum must be done on a more complicated
complex path for avoiding the singularities of the integrands.
The technical aspects of this procedure are discussed
in detail in ref. \cite{forthcoming}.

Once the subtracted $1/N$ graphs are calculated, they can
be used for deriving physical amplitudes. Here we will
consider the Higgs resonance shape in the scattering process
$f\bar{f} \rightarrow H \rightarrow WW$. This process is for instance
relevant for direct searches at a possible muon collider. 
Also it is related to the corrections of enhanced electroweak
strength to the Higgs production mechanism by gluon 
fusion at hadron colliders \cite{gluefusion}. 

At NLO in the $1/N$ expansion, by including the relevant $1/N$
graphs of all loop-orders and the corresponding counterterms
derived from the Lagrangian of eq. 1, one obtains the following
nonperturbative expression for the 
$f\bar{f} \rightarrow H \rightarrow WW$ scattering amplitude 
(the overall factor from the tree level Yukawa coupling not included):

\begin{eqnarray}
{\cal M}_{WW} & = &
\frac{m^2(s)}{\sqrt{N} v}
\frac{1 - \frac{1}{N} f_2 }{s - m^2(s) \left[ 1 - \frac{1}{N} f_1(s)  \right] }
  \nonumber  \\
f_1(s)  & = &
           \frac{m^2(s)}{v^2} \, \hat{\alpha}(s) 
       + 2 \hat{\gamma}(s)
       +   \frac{v^2}{m^2(s)} \left[ \hat{\beta}(s)
                               - 2 \frac{s-m^2(s)}{v^2} \left( \delta Z_{\sigma} - \delta Z_{\pi} \right)
                              \right]
  \nonumber  \\
f_2(s)  & = &
           \frac{m^2(s)}{v^2} \, \hat{\alpha}(s) 
       +   \hat{\gamma}(s)
       -   \hat{\phi}(s)
       -   \frac{v^2}{m^2(s)} \hat{\eta}(s)
\end{eqnarray}

Note that this expression is renormalization scheme independent.
Here we used the notation $\hat{\eta}(s)=\hat{E}_1(s)$ and
$\hat{\phi}(s)=\hat{F}_1(s)+\hat{F}_2(s)+\hat{F}_3(s)$. Similarly,
$\hat{\alpha}(s)$, $\hat{\beta}(s)$ and $\hat{\gamma}(s)$
are the subtracted two-point functions of the model,
and $\delta Z_{\sigma}$ and $\delta Z_{\pi}$ are the wave function
renormalization constants of order $1/N$ of the Higgs and Goldstone fields, 
which were defined and calculated in ref. \cite{oneovn}. $m^2(s)$ is the
LO self-energy (see the notations of ref. \cite{oneovn}),
and $\sqrt{N}v=246$ GeV is the vacuum expectation value of the Higgs field.

\begin{figure}[t]
\hspace{1.5cm}
    \epsfxsize = 15cm
    \epsffile{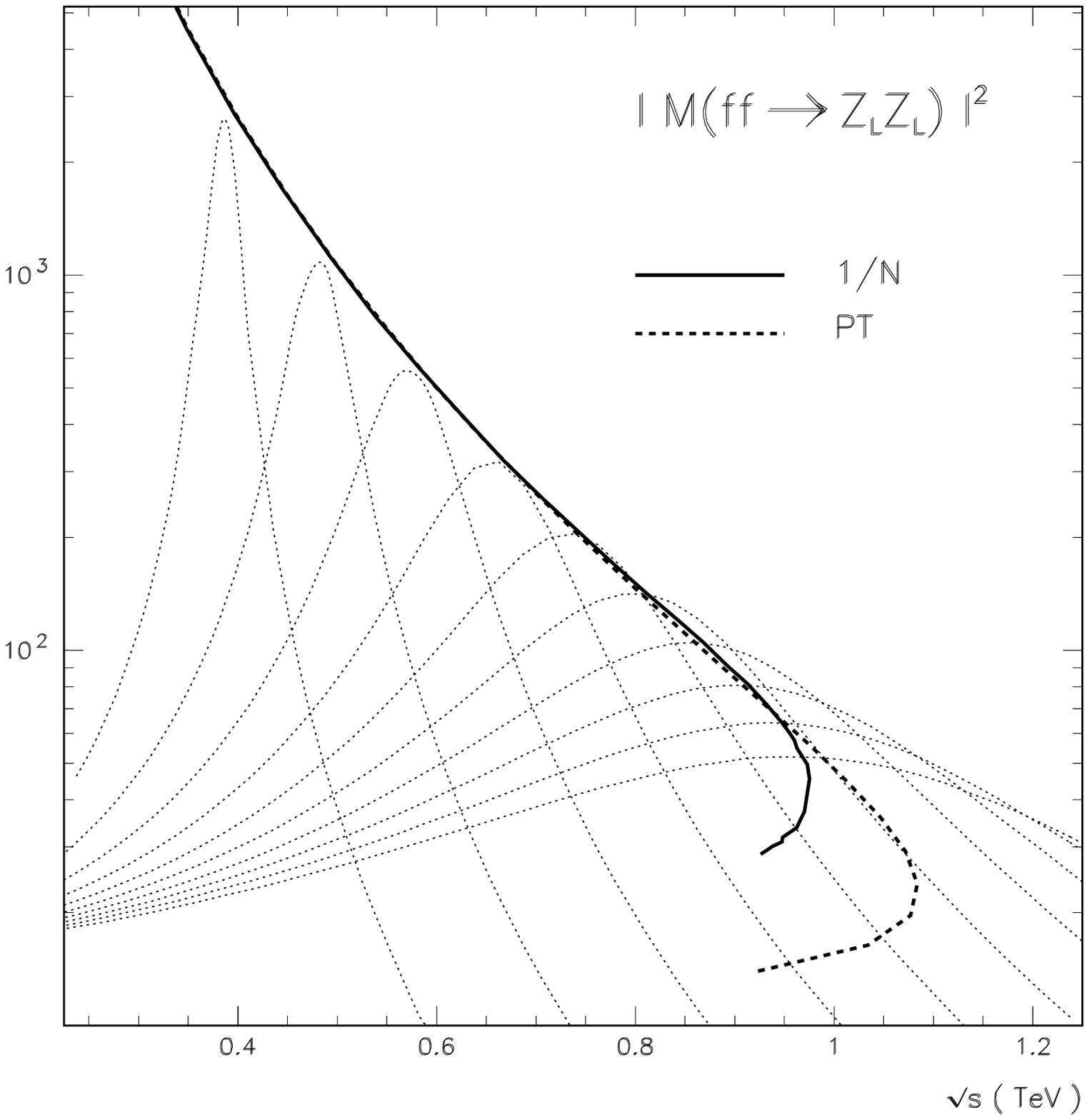}
\caption{{\em The line shape of the Higgs resonance in the scattering
              process
              $f\bar{f} \rightarrow H \rightarrow ZZ, WW$ for different
              values of the quartic coupling. The solid
              line indicates the position of the maxima of the resonances
              of the nonperturbative $1/N$ expansion when the coupling
              is increased. The dotted line corresponds to the 
              two-loop perturbative scattering amplitude.}}
\end{figure}

We plot in figure 2 a set of nonperturbative line shapes of the 
Higgs resonance as it appears in this scattering process. We also
plot the location of the peaks of these line shapes. For
comparison we show also the corresponding location of the peaks
in perturbation theory, calculated in two-loop order
\cite{mucollider,gluefusion}. Just
as in the case of the 
$f\bar{f} \rightarrow H \rightarrow f^{\prime}\bar{f^{\prime}}$ 
scattering process \cite{oneovn}, the NLO $1/N$ solution agrees well
with two-loop perturbation theory at weak coupling. As
the coupling increases, a saturation effect sets in \cite{oneovn},
and the width of the resonance increases without the position
of the peak increasing at the same time. Nonperturbatively,
the saturation value of the position of the peak is at
about 980 GeV. This is not far from the 930 GeV value found
in the fermion scattering process \cite{oneovn}. 
Of course, the shape of the 
resonance is process dependent, as opposed to the position 
of the complex pole of the Higgs particle, which is universal.

To conclude, we calculated nonperturbatively the three-point functions
of the scalar sector of the standard model by means of
a $1/N$ expansion at next-to-leading order. Combining the three-point
function with the already available two-point functions, we derived
the nonperturbative amplitude of the scattering process 
$f\bar{f} \rightarrow H \rightarrow ZZ, WW$. Similarly to the
already known 
$f\bar{f} \rightarrow H \rightarrow f^{\prime}\bar{f^{\prime}}$ 
scattering, the NLO $1/N$ solution agrees very well with two-loop
perturbation theory at weak coupling. At strong coupling we confirm
the existence of a Higgs mass saturation effect. In this process
the saturation value is about 980 GeV. This is comparable with the
value of 930 GeV obtained from fermion scattering.



\end{document}